\documentclass[showpacs,amsmath,amssymb,superscriptaddress,aps,raggedbottom,twoside,twocolumn]{revtex4}

\usepackage[hidelinks]{hyperref}
\usepackage{graphicx}
\usepackage{color}

\usepackage{xargs}                      % Use more than one optional parameter in a new commands
\usepackage[pdftex,dvipsnames]{xcolor}  % Coloured text etc.

\newcommand{\vv}[1]{\mathbf{#1}}

\newcommand{\vFermi}{v_{\mathrm{F}}}

\newcommand{\atanh}[1]{\,\mathrm{arctanh}\left(#1\right)\,}

\newcommand{\gap}{\Delta}

\newcommand{\TE}{^{\mathrm{TE}}}
\newcommand{\TM}{^{\mathrm{TM}}}

\newcommand{\dd}{d}

\newcommand{\ee}[1]{\mathrm{e}^{#1}}

\usepackage[normalem]{ulem}

\begin{document}

\title{Determining graphene's induced band gap with magnetic and electric emitters}

\author{Julia F.M. Werra}
\email{jwerra@physik.hu-berlin.de}
\affiliation{
  Humboldt-Universit\"{a}t zu Berlin, Institut f\"{u}r Physik, AG Theoretische Optik \& Photonik, Newtonstra{\ss}e 15,
  12489 Berlin, Germany 
}
\author{Peter Kr\"uger}
\affiliation{
  Midlands Ultracold Atom Research Centre, School of Physics and Astronomy, University of Nottingham, Nottingham NG7
2RD, United Kingdom}
\author{Kurt Busch}
\affiliation{
  Humboldt-Universit\"{a}t zu Berlin, Institut f\"{u}r Physik, AG Theoretische Optik \& Photonik, Newtonstra{\ss}e 15,
  12489 Berlin, Germany
}
\affiliation{
  Max-Born-Institut, Max-Born-Stra{\ss}e 2A, 12489 Berlin, Germany
}
\author{Francesco Intravaia}
\affiliation{
  Max-Born-Institut, Max-Born-Stra{\ss}e 2A, 12489 Berlin, Germany
}
\date{\today}

\pacs{81.05.ue,78.67.Wj,73.20.Mf,13.40.Hq}

\begin{abstract}
We present numerical and analytical results for the lifetime of emitters in close 
proximity to graphene sheets. Specifically, we analyze the contributions from 
different physical channels that participate in the decay processes. Our results 
demonstrate that measuring the emitters' decay rates provides an efficient route 
for sensing graphene's optoelectronic properties, notably the existence and size 
of a potential band gap in its electronic bandstructure.
 \end{abstract}

\maketitle

Driven by its successful isolation, graphene has not stopped fascinating the research
community.  Although, this allotropic form of carbon had been theoretically 
investigated for decades, experimental access to graphene has offered new perspectives 
as well as novel  directions for fundamental research and technological applications 
\cite{Geim07,Geim09}.
Graphene's exotic properties \cite{Castro-Neto09} have lead to the investigation of 
a wide range of phenomena such as ballistic transport \cite{Bae13}, 
the quantum Hall effect \cite{Geim07,Novoselov07}, and 
thermal \cite{Balandin11} 
as well as electrical conductivity \cite{Wunsch_2006,Guo11}. 
Developing a detailed understanding, followed by appropriate engineering of these 
properties, lies at the heart of future graphene-based technologies. For this, an 
accurate determination of graphene's properties in realistic experimental settings 
and the detailed validation of various theoretical models (cf. Ref~\cite{Wunsch_2006,
Hwang_2007,Bordag_2014,Bordag_2015}) is indispensable. Promising designs where the 
semi-metal will play an important role, aim at combining condensed-matter  with 
atomic systems. Such hybrid devices are geared towards reaping the best of the two 
worlds for advanced high-performance devices.

\begin{figure}[t]
\centering
\includegraphics[width=8cm]{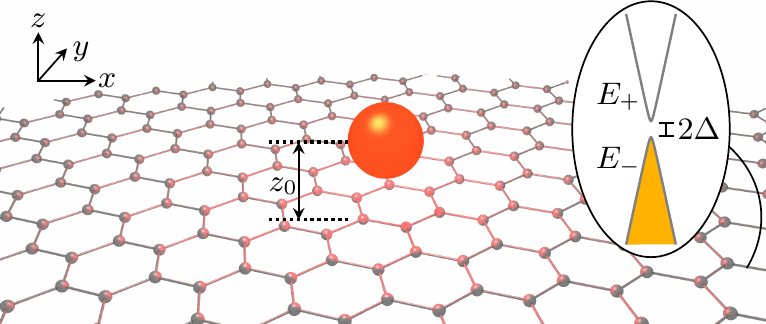}
\caption{Schematic of the physical situation considered in this work. An emitter 
         (red sphere) is positioned at distance $z_0$ from a graphene sheet. 
	 Graphene's bandstructure is approximated by $E_{\pm}=\pm\sqrt{\gap^2
	 +\vFermi^2k^2}$ (see inset) and the chemical potential is chosen as $\mu=0$
         (yellow: filled band).}
\label{fig:grapheneEmitterSketch}
\end{figure}

In this work, we demonstrate how the high degree of control and accuracy available 
in quantum systems like cold atoms and Si- and NV-centers in nano-diamonds, can be 
employed for detailed investigations of graphene's optoelectronic properties 
\cite{Gomez-Santos_2011,Nikitin_2011,Tielrooij15,Kort_2015}.
Specifically, we focus on modifications in the life times of emitters held in close 
proximity of graphene layers and show that these allow for direct experimental 
access to features like band gaps as well as plasmons and/or plasmon-like 
resonances.
In graphene, a band gap $\Delta$ 
(cf. Fig.~\ref{fig:grapheneEmitterSketch}) is created
(i) when the atomically thin material is deposited on a substrate 
  \cite{Giovannetti07,Jung_2015},
(ii) when strain is applied, 
(iii) when impurities are present, and
(iv) in cases where graphene bilayers instead of a single layer are considered. 
Values for $\Delta$ of the order of tens of meV have been predicted 
\cite{Giovannetti07,Jung_2015}, thus triggering corresponding experimental 
investigations. These band gaps and the features connected with them are 
still the subject of discussions \cite{Bai_2015,Kumar_2015}
so that reliable experimental means for their analysis are highly desirable.  

For planar geometries the decay rate of an emitter is a functional of the system's 
optical scattering coefficients. We model a monoatomic graphene layer in terms of 
a 2+1-dimensional Dirac fluid \cite{Fialkovsky_2011,Chaichian_2012,Bordag_2014}
and embed it in a non dispersive and non dissipative dielectric medium with 
permittivity $\varepsilon_{m}$. As a result, the graphene layer is characterized by 
an induced band gap and a chemical potential $\mu=0$ 
(cf. Fig.~\ref{fig:grapheneEmitterSketch}) while the corresponding electromagnetic 
reflection coefficients for transverse magnetic (TM) and transverse electric (TE) 
waves are \cite{Fialkovsky_2011,Chaichian_2012}
\begin{align}
\label{rcoeff}
 r\TM=-\frac{\alpha\Phi(y)}{y\varepsilon_m/\kappa_m - \alpha\Phi(y)}\,,\,\,
 r\TE=-\frac{\alpha\Phi(y)}{\kappa_m+\alpha\Phi(y)}\,.
\end{align}
where $\alpha=137^{-1}$ is the fine structure constant and
\begin{align}
\Phi(y)=1-\left(\sqrt{y}+1/\sqrt{y} \right) \atanh{\sqrt{y}}\,,
\end{align}
with $y =\omega^2-\vFermi^2k^2$. Further, $\kappa_m=\sqrt{k^2-\varepsilon_m\omega^2}$ and
$k=\sqrt{k_x^{2}+k_y^{2}}$ denote, respectively, the moduli of the out-of-plane and in-plane 
wave vectors in the dielectric medium. In addition, we use dimensionless variables, which amounts to the replacements $\hbar\omega/2\gap\to \omega$, $\hbar c k/2\gap\to k$, and 
$\vFermi/c\to \vFermi$ ($\approx 300^{-1}$ for graphene).
Life time modifications are usually associated with the strength of scattering 
processes. Owing to its minute thickness (few \AA), the optical response of a single 
graphene layer is rather small ($\sim2\%$ reflection \cite{Nair_2008}). Thus, for 
emitters near a graphene layer, small life time modifications 
might naively be expected. However, graphene's exotic properties introduce additional 
features that affect the emitters' dynamics, such as TE plasmons and single (SPE)- and 
multiple (MPE)-particle excitations.

Different frequencies are associated with the different physical processes: 
propagating fields occur for $0\le k<\omega\sqrt{\varepsilon_{m}}$ and evanescent 
fields are characterized by $k>\omega\sqrt{\varepsilon_{m}}$. Further, we identify 
another regime where $k<\sqrt{\omega^{2}-1}/\vFermi$, which only exists if 
$\omega>1$,i.e., if the radiation frequency exceeds that associated with the band gap. 
In this regime, the 2+1-dimensional Dirac fluid model of graphene features the 
creation of electron-hole pairs. In the propagating regime, the scattering process in 
graphene systems is very similar to that in ordinary thin films. This similarity, 
however, already breaks down for evanescent waves, for which the scattering process 
is associated with surface plasmons or plasmon-like phenomena: While in ordinary 
materials these resonances are usually present only in TM polarization, graphene is 
known for admitting such excitations in both TM and TE polarization 
\cite{Mikhailov_2007,Bordag_2014,Stauber_2014,Bordag_2015}. 
TM polarized surface plasmons are associated with charge density oscillations and are
dominated by the electric field. Conversely, TE plasmons result from resonances in
the motion of the current density so that they are dominated by the magnetic field.
Mathematically, these phenomena are related to divergences of the scattering 
coefficients and in our case they can be investigated by analyzing the poles of Eqs.~
\eqref{rcoeff}. In our model TM plasmons do not occur, while the TE plasmon's 
dispersion relation reads as
\begin{equation}
\label{TEDispersion}
\begin{cases}
\omega[y]=\sqrt{y+\vFermi^2 k\left[y\right]^2}\\
 k[y]=\omega_{\mathrm{g}}\sqrt{\alpha^2\Phi\left(y\right)^2+\varepsilon_m y}~, 
\end{cases}
\end{equation}
where $\omega_{\mathrm{g}}=1/\sqrt{1-\varepsilon_{m}\vFermi^2}$.
This agrees well with previous numerical results for vacuum ($\varepsilon_m=1$) 
\cite{Bordag_2014}.
\begin{figure}[h!]
\centering
\includegraphics[width=8.5cm]{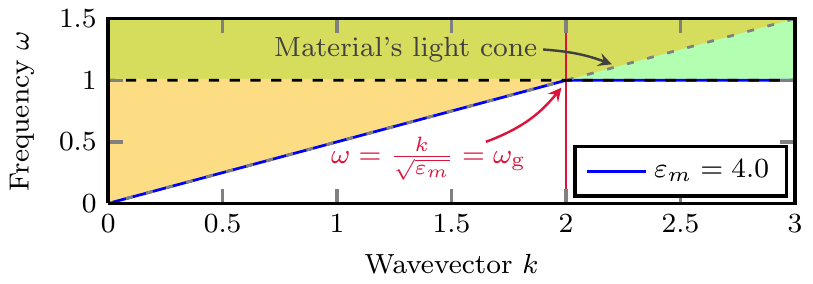}
\caption{Dispersion relation of the TE surface plasmon (blue line) for a graphene 
         layer embedded in a dispersionless dielectric material ($\varepsilon_{m} 
	 = 4.0$). The colored areas delineate the wave vector regions corresponding 
	 to different decay channels: propagating waves (yellow), evanescent waves 
	 (white), and single particle excitations (green).}
\label{fig:TEplasmonMode_vacuum}
\end{figure} 
Albeit difficult to discern in Fig.~\ref{fig:TEplasmonMode_vacuum}, Eq.~
\eqref{TEDispersion} indicates that the TE plasmon's dispersion relation lies 
exclusively in the evanescent region and stays outside of the single-particle 
excitation region (SPE) \cite{TE-Plasmon-Paper-auf-ArXiV}. Two distinct 
characterstics become apparent: For low frequencies ($\omega<\omega_{\mathrm{g}}$), 
the dispersion curve lies close to but below the medium's light cone; For large 
frequencies ($\omega>\omega_{\mathrm{g}}$), the properties of the TE plasmon's 
dispersion do not depend on the embedding dielectric but are solely determined by 
graphene itself. 

With respect to the processes described above, the total decay rate of an emitter 
with dipole operator $\hat{\vv{d}}$ can be written as $\gamma/\gamma_{0}=1+
\mathcal{L}[(d_{\|}^2/|\vv d|^2)\Gamma^{\|}+(d_{\perp}^2/|\vv d|^2)\Gamma^{\bot}]$ 
where $\gamma_0$ is the decay rate in a homogeneous dielectric without graphene. The 
factor $\mathcal{L}$ indicates the usually frequency-dependent local field 
correction one has to take into account to correctly describe the dynamics of an 
emitter embedded in a dielectric ($\mathcal{L}=1$ for $\epsilon_{m}=1$) 
\cite{Glauber91,Intravaia15a}. For simplicity, we will not dwell on this issue and 
instead refer readers to the literature for further information \cite{Glauber91,Intravaia15a,
Scheel99,Vries98}. The functions $\Gamma^{\|,\bot}$ are related to the matrix 
elements of the orthogonal $d_{\perp}$ and parallel $d_{\|}$ components of the dipole 
with respect to the graphene layer ($|\vv d|^2=d_{\|}^2+d_{\|}^2$). In turn, each of
these two contributions is the result of the three processes discussed above. 
Consequently, we have a the radiative term $\Gamma_{\mathrm{r}}$, which originates 
from the propagating region (including the radiative part of the SPE region), the 
contribution of the (non-radiative) SPE region $\Gamma_{\mathrm{SPE}}$, and the 
non-radiative contribution given by plasmonic excitations $\Gamma_{\mathrm{p}}$.

\begin{figure*}
\centering
\includegraphics[width=17cm]{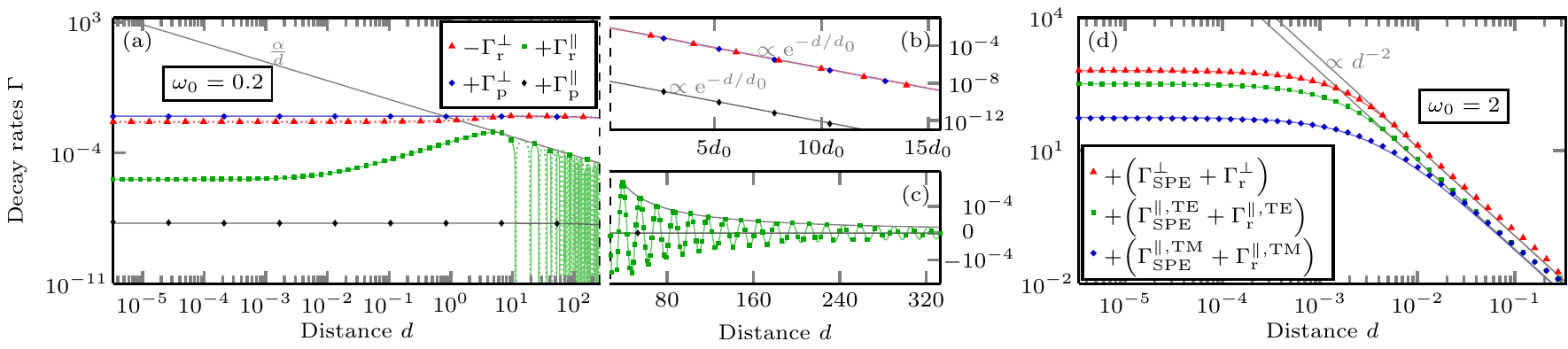}
\vspace{-0.2cm}
\caption{Distance behavior of the different contributions to a magnetic emitter's 
         decay rate for an emitter above a free-standing graphene layer ($\epsilon_{m}=1$). Panels 
	 (a)-(c) and (d) display results for transistion frequencies well below 
         ($\omega_0 = 0.2$) and well above ($\omega_0 = 2.0$) the band gap of 
	 graphene, respectively. Symbols correspond to complete numerical solutions
         and lines to approximate analytical solutions (see text for detail). Panels
         (b) and (c) represent continuations of the x axis shown in panel (a) and 
	 panel (c) is plotted on a linear scale.
         Note that the results are plotted on a logarithmic scale so that 
	 contributions leading to an enhancement (+) or a suppression (-) of the 
	 decay rate are indicated by corresponding signs in the inset of panels (a) 
	 and (d). Further, for distances $d>10$ the (for small distances strictly 
	 positive) contributions from $\Gamma_r^\parallel$ oscillate around zero as 
	 depicted in panel (c).
         }
\label{fig:lifetime_distDep_Total}
\end{figure*}

In order to analyze the above terms in more detail, we will first discuss the case of 
magnetic decay keeping in mind that a magnetic emitter ought to be more sensitive 
to the magnetic field associated with plasmonic TE resonances. The emitter has a 
transition frequency $\omega_0$ and is located at $z=z_{0}>0$ above the graphene 
layer at $z=0$ (see Fig.~\ref{fig:grapheneEmitterSketch}). Within second-order 
perturbation theory \cite{Henkel_1999,Novotny_2012} the modification of the decay 
rate can be written as
\begin{subequations}
\label{eq:magneticDecRate}
\begin{align}
\hspace*{-0.2cm}
\Gamma^{\|}&=\frac{3}{4}
\int\limits_{-\infty}^{\omega_{0}^{2}}
\hspace{-0.1cm}
\dd y\,\mathrm{Im}\left[
\frac{\frac{r\TM}{K_{s}[y]}+ \frac{K_{s}[y] r\TE}{k_{0}^{2}\vFermi^{2}}}{2k_{0}\vFermi} \,
\exp \left(- 2d \frac{K_{s}[y]}{\vFermi} \right)   \right],
\\
\Gamma^{\bot}&=\frac{3}{2}
\int\limits_{-\infty}^{\omega_{0}^{2}}\hspace{-0.1cm}\dd y\,\mathrm{Im}\left[
              \frac{k_{s}[y]^{2}}{K_{s}[y]}
              \frac{r\TE}{2k_{0}^{3}\vFermi^{3}} 
              \exp \left(- 2d \frac{K_{s}[y]}{\vFermi} \right)  \right].
  \end{align}
\end{subequations}
Here, $k_{0}=\omega_0\sqrt{\varepsilon_{m}}$, $d=2 z_{0}\Delta/\hbar c$. We have also 
defined $k_{s}[y]=\sqrt{\omega_{0}^{2}-y}$ and 
$K_{s}[y]\equiv\vFermi\kappa[y]=\sqrt{\omega_{0}^{2}/\omega_{\mathrm{g}}^{2}-y}$.
In Eqs. \eqref{eq:magneticDecRate}, the evanescent contribution is associated with 
the range $-\infty\le y\le (\omega_{0}/\omega_{\mathrm{g}})^{2}$, while the 
$(\omega_{0}/\omega_{\mathrm{g}})^{2}\le y\le \omega_{0}^{2}(<1)$ corresponds to the 
propagating region. The SPE range corresponds to $1<y<\omega_0^2$.

We first consider the contribution to the decay rate from the evanescent range, 
imputable only to the resonance in the reflection coefficients. In view of the 
above discussion of the dispersion relation, Eqs.~\eqref{TEDispersion}, this 
contribution features two different regimes. For $\omega_{0}<\omega_{\mathrm{g}}$, 
i.e., when the dispersion curve is very close to the light cone, the resonance is 
located at $ y_{\mathrm{p}} \approx (\omega_{0}/\omega_{\mathrm{g}})^{2}
[1-\left(4\alpha\vFermi/3\right)^2(\omega_{0}/\omega_{\mathrm{g}})^{2}]$. 
The leading terms of Eqs.~\eqref{eq:magneticDecRate} are then
\begin{equation}
\Gamma^{\|}_{\mathrm{p}} \approx  \frac{16\alpha^3\pi}{9\varepsilon_m^{3/2}}
                                  \frac{\omega_0^3}{\omega_{\mathrm{g}}^3 } 
                                  \exp (-d/d_0),
\,\,
\Gamma^{\bot}_{\mathrm{p}} \approx \frac{2\pi\alpha}{\sqrt{\varepsilon_m}}
                                   \frac{\omega_0}{\omega_{\mathrm{g}}^2} 
                                   \exp (-d/d_0)~.
\label{Pl-omegaSchi}
\end{equation}
Given the rather large characteristic decay length 
$d_0=[3\epsilon_{m}\omega_{\mathrm{g}}^{2}/(8\alpha)]k_{0}^{-2}$, these contribitions 
exhibit weak distance-dependencies for experimentally relevant emitter-graphene 
separations of a few microns. For $\omega_{0}>\omega_{\mathrm{g}}$, the resonance 
is instead located close to the boundary of the SPE region, 
$y_{\mathrm{p}}\approx1-2\exp[-(1+K_{s}[1]/(\alpha\vFermi)) ]$ and we obtain
\begin{subequations}
\begin{align}
\Gamma^{\|}_{\mathrm{p}} & \approx 
          \frac{3\pi K_{s}[1]\,\ee{-\left(1+\frac {K_{s}[1]}{\alpha\vFermi} 
	      \right)}}{2\alpha\vFermi^4k_0^3} 
          \exp(-2d\frac{K_{s}[1]}{\vFermi }) ,\\
\Gamma^{\bot}_{\mathrm{p}} & \approx
          \frac{3\pi k_{s}[1] \ee{-\left(1+\frac{K_{s}[1]}{\alpha\vFermi} 
	      \right)}}{\alpha\vFermi^4k_0^3} 
          \exp(-2d\frac {K_{s}[1]}{\vFermi}) \, .
\end{align}
\end{subequations}
Due to the small values of $\vFermi$ and $\alpha$, the above terms are strongly 
suppressed in graphene unless $K_{s}[1]\sim 0$, which only occurs when 
$\omega_{0}\sim \omega_{\mathrm{g}}\gtrsim 1$. 

For the same parameters, the propagating regime corresponds to a rather small 
integration range in Eqs.~\eqref{eq:magneticDecRate}. Therefore, the integrands 
can be expanded around $y=\omega^{2}_{0}$ and after some rearrangements we obtain
\begin{subequations}
\label{eq:scatteringMagneticOmSmallerZero}
\begin{align}
 \Gamma^{\|}_{\mathrm{r}}&\approx\frac{\alpha(\varepsilon_m+1)}{2\varepsilon_m d}
   \left[\frac{4\alpha k_0(\varepsilon_m+3)}{9\varepsilon_m(\varepsilon_m+1)}
     \sin{\left(2k_0d\right)}\right.\nonumber\\
   &\qquad\qquad\left.+\frac{\sin{\left(2 k_0 d\right)}}{2k_0d}-\cos{\left(2 k_0 d\right)}
   \right], \\
 \Gamma^{\perp}_{\mathrm{r}}&\approx-\frac{\omega_{g}^{2}d}{2d_{0}}\int
   \limits_0^{2k_0d }d\zeta\frac{\left[1-\left(\frac{\zeta}{2k_0d}\right)^{2}
   \right]\left[\frac{\zeta\sin{\left(\zeta\right)}}{2k_{0}d}
   +\frac{\omega_{g}^{2}\cos{\left(\zeta\right)}}{2k_{0}d_{0}} \right]}
    {\left(\frac{\omega_{g}^{2} d}{d_{0}}\right)^{2}+\zeta^2}\nonumber\\ 
   &\stackrel{d\ll k_0^{-1}}{\approx\quad}-\frac{\pi \alpha}
    {\sqrt{\varepsilon_m}}\omega_0\left(1+\frac{8k_0d}{3\pi}\right)~.
\end{align}
\end{subequations}
Interestingly, because of the overall minus sign of $\Gamma^{\perp}_{\mathrm{r}}$, 
this contribution tends to increase the emitter's life time, suppressing the decay 
process relative to $\gamma_{0}$. In addition, since 
$k_{0}^{-1}\ll d_{0}/\omega_{g}^{2}$, due to the dephasing between the propagating 
waves,  $\Gamma^{\perp}_{\mathrm{r}}$ exponentially decays for distances 
$d\gtrsim d_{0}/\omega_{g}^{2}$. It follows a behavior similar to the TE plasmon but 
with characteristic decay length $d_{0}/\omega_{g}^{2}$. Therefore, since 
$\omega_{g}\sim 1$, $\Gamma^{\perp}_{\mathrm{r}}$ is almost exactly canceled by 
$\Gamma_{\rm p}^{\bot}$ (see Fig.~\ref{fig:lifetime_distDep_Total}(b)). For even 
larger distances ($d\gg d_{0}/\omega_{g}^{2}$, not shown), due to the interference 
between incoming and scattered waves, $\Gamma_{\rm r}^{\bot}$ oscillates in space 
like $\Gamma_{\rm r}^{\|}$ with a frequency $2k_{0}$
(see Fig.~\ref{fig:lifetime_distDep_Total}(c)).

\begin{figure*}
\centering
\vspace{-0.2cm}
\includegraphics[width=17cm]{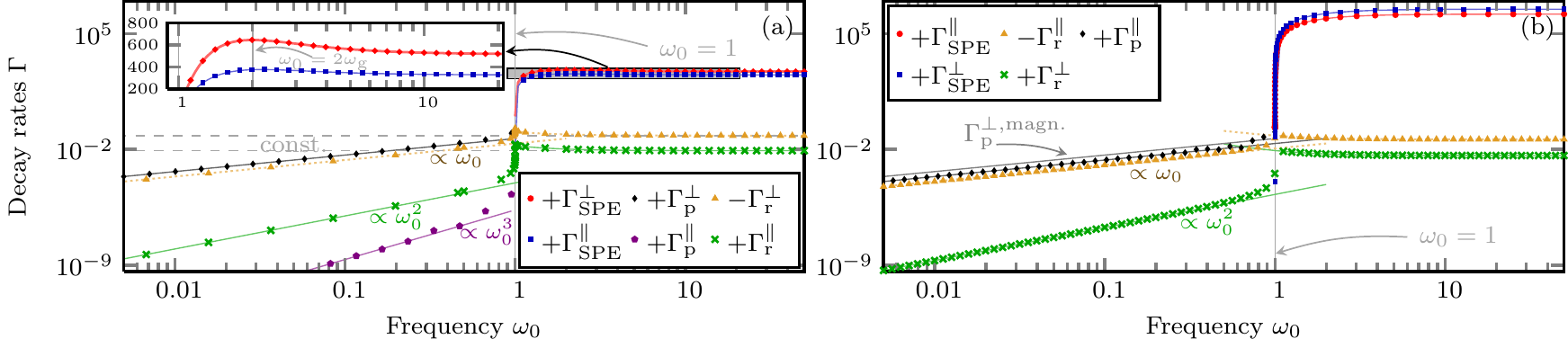}
\caption{Orthogonal and parallel decay rates of an emitter situated at 
         $d=(3\cdot10^{8})^{-1}$ above a graphene layer suspended in air. The lines 
         represent the analytical approximations discussed in the text while the dots 
	 represent numerical results. Panel~(a): Results for a \emph{magnetic 
	 emitter}, Panel~(b): Results for an \emph{electric emitter}. Note that the 
	 results are plotted on a logarithmic scale so that contributions which lead 
         to an enhancement (+) or a suppression (-) of the decay rate are signs as 
	 indicated by corresponding signs in the insets of panels (a) and (b).}
\label{fig:lifetime_freqDep_Total}
\end{figure*}

Finally, we consider the modification of the decay rate that stems from the SPE 
region. This contribution only occurs when the emitter's transition frequency 
becomes larger than the electronic band gap ($\omega_{0}>1$). Although, the total SPE 
region includes both evanescent and propagating contributions, the non-radiative part 
dominates at short distances and, as in the previous case, is almost constant for 
$d\ll k_{0}^{-1}$. Again, since $\alpha,\vFermi \ll1$,  in this limit we can write 
\begin{align}
\label{eq:SPEorth}
\Gamma^{\perp}_{\mathrm{SPE}}
&\approx\frac{\alpha\pi}{4\vFermi^2\varepsilon_1^{3/2}\omega_{\mathrm{g}}^3}
  \left[1+3\left(\frac{\omega_{\mathrm{g}}} {\omega_{0}}\right)^{2}-4
    \left(\frac{\omega_{\mathrm{g}}} {\omega_{0}}\right)^{3} 
  \right]\,,
\end{align}
This demonstrates that $\Gamma^{\perp}_{\mathrm{SPE}}$ varies non-monotonously with 
frequency and exhibits a maximum for $\omega_{0}=2\omega_{\mathrm{g}}$, where it 
takes the value $\Gamma^{\perp}_{\mathrm{SPE}}\approx 645$. At intermediate 
distances, the total (evanescent and propagating) SPE contribution decays as a power 
law, $\Gamma^{\perp}_{\rm SPE}+\Gamma^{\perp}_{\rm r}
            \approx
            2 ( \Gamma^{\|}_{\rm SPE}+\Gamma^{\|}_{\rm r})
            \approx
            \alpha\pi\omega_{\mathrm{g}}(\omega_0^{-2} +1)(6\omega_0d)^{-2}$ 
(see Fig.~\ref{fig:lifetime_distDep_Total}(d)). For $d\gg k_{0}^{-1}$, the 
propagating waves induce once again spatial oscillations with frequency $2k_{0}$ 
(not shown).

In Fig.~\ref{fig:lifetime_freqDep_Total}(a) we present the frequency dependence 
of all the above-discussed contributions to the decay rate at a fixed distance 
$d=(3\cdot10^{8})^{-1}$ from the graphene layer (corresponding to 
$z_0=1\,\mu\mathrm{m}$ for an emitter with transition frequency of $1\,\mathrm{MHz}$). 
As discussed above, for emitters with transition frequencies smaller than 
graphene's electronic band gap ($\omega_{0}<1$), the two main decay channels 
are the TE plasmonic resonance and the radiative decay. Their relative importance 
differs, depending on the spatial orientation of the dipole-matrix elements. 
We see that in $\Gamma^{\perp}$ the plasmonic TE resonance provides an enhancement 
while the radiative contribution leads to a suppression. Also, 
$\Gamma^{\perp}_{\rm p}\approx -2 \Gamma^{\perp}_{\rm r}$ over a very large range 
of frequencies. For $\Gamma^{\|}$, the radiative contribution dominates and leads 
to an enhancement of the decay rate. In this case, the plasmonic TE resonance, 
due to its proportionality to $\omega_{0}^{3}$, represents a subleading contribution. 
For $\omega_{0}>1$, the dominant contribution for both $\Gamma^{\perp}$ and 
$\Gamma^{\|}$ stems from the SPE contribution (see 
Fig.~\ref{fig:lifetime_freqDep_Total}(a), inset) and leads to an enhancement of the 
decay rate by three orders of magnitude. Note, that the increase of the decay rate 
occurs quite abruptly as the frequency of the emitter moves across the band gap and, 
for larger frequencies, takes on a weakly frequency-dependent value around 
$\alpha\pi/(4 \vFermi \omega_{\mathrm{g}}^{3} \varepsilon_{m}^{3/2})\approx10^{3}$. 
In both $\Gamma^{\|,\perp}$, the TE contributions are dominant and lead to the 
non-monotonic behavior discussed above. 

Most of the above-described characteristics also qualitatively apply to the case 
of an electric dipole emitter (see Fig.~\ref{fig:lifetime_freqDep_Total}(b)). 
Indeed, the relevant expressions can be easily obtained by swapping the reflection 
coefficients in Eqs.~\eqref{eq:magneticDecRate} \cite{Henkel_1999,Intravaia11}. 
For brevity we will only mention that, as a consequence of the replacement 
$r\TM\leftrightarrow r\TE$, some features are found in $\Gamma^{\|}$ instead of 
$\Gamma^{\bot}$. Curiously, for $\omega_{0}<\omega_{g}$, due to the proximity 
of the TE plasmon dispersion relation to the light cone, its contribution to the 
decay rate is of the same order of magnitude for both emitters and for all distances, 
i.e., $\Gamma_{\mathrm{p}}^{\mathrm{el.}}\approx2\Gamma_{\mathrm{p}}^{\mathrm{magn.}}/\sqrt{\varepsilon_m}$. 
More importantly, the SPE channel still provides a large enhancement of the decay 
rate for $\omega_{0}>1$, featuring again a quite abrupt jump for frequencies near 
the electronic band gap of graphene. However, for the electric emitter both 
$\Gamma^{\|,\bot}$ exhibit a monotonous frequency dependence.

In conclusion, the above results suggest atomic or atom-like emitters as sensitive 
quantum probes to determine the physical properties of graphene and, in particular, 
to investigate a band gap in its electronic bandstructure. Using these systems allows 
for an accurate analysis of this quantity, especially in complex (but relevant for graphene-based technologies) situations 
where it is no longer spatially homogenous: This occurs, e.g., when the sheet (i) is exposed to mechanical stress \cite{Zhu15}, (ii)
is positioned on an inhomogeneous substrate or (iii) absorbs impurities (in a controlled \cite{Elias09} or uncontrolled fashion).
In our approach, the emitter non-invasively probes graphene's properties 
in different physical regimes, enabling experimental investigation 
of unusual graphene properties such as TE surface resonances (see also \cite{Gomez-Santos_2011,
Nikitin_2011}) and 
providing results complementary to those accessible when using other procedures.
In addition, the possibility to engineer different internal quantum states of the emitter and study their 
lifetimes can also offer new opportunities which are presently not accessible with other techniques.
As a concrete experimental approach, we suggest to extending the known use of microtrapped 
Bose-Einstein condensates \cite{Wildermuth_2005,Wildermuth_2006} to map the local 
band gap structure of graphene sheets with micron resolution. One would 
detect the splin flip rate by measuring the spatially dependent spin population after 
a known time since its preparation as a spin-polarized gas. For enhanced sensitivity, 
it will be advantageous to employ an optical dipole trap, ideally configured as a 
light sheet, tuned to a frequency below the main atomic transition. Fluorescence 
imaging following selective resonant excitation of the emitter decay target state 
will enable the measurement of even very slow decay rates down to a few events per 
time across the ensemble of typically $10^5$ atoms. 
The high temporal resolution of this technique can offer an important advantage 
in analyzing the different (relatively slow) processes cited above.

In addition to atomic quantum 
gases other very well suited candidates are Si- and NV-centers in nano-diamonds.
They do not only show tunable magnetic and electric transitions from the MHz to the 
THz frequency range but also simultaneously allow for high position resolution 
\cite{Schell_2014}. Small band gaps can be investigated by cooling the system to the 
mK regime, such that magnetically tunable Zeeman \cite{Amsuss_2011} or hyperfine 
transitions \cite{Tkalcec_2014} can be utilized. 
Our work can open additional pathways to enhance 
the fundamental understanding of the validity of different graphene models 
\cite{Stauber_2014,Roldan_2013,Brida_2013} and also provides relevant information for 
realistic applications and new designs of interest, e.g., in atom-chip research \cite{Folman02,Fortagh07,Sinuco-Leon11}. 
Indeed, this material with its intrinsic, room-temperature quantum properties 
\cite{Geim09,Novoselov07,Tombros_2007} has been deemed as a particularly interesting 
addition to these systems in order to proceed further on the road to quantum computing 
\cite{Trauzettel_2007,Guo_2009}.
\paragraph*{Acknowledgements.}
%\begin{acknowledgements}
We thank Ch.~Koller, M.~T.~Greenaway and T.~M.~Fromhold for stimulating discussions. We acknowledge 
support by the Deutsche Forschungsgemeinschaft (DFG) through the Collaborative Research 
Center (CRC) 951 ``Hybrid Inoganic/Organic Systems for Optoelectronics (HIOS)" within Project No. B10.
P.K. acknowledges support from EPSRC (grant EP/K03460X). F.I. further acknowledges financial support 
from the EU through the Career Integration Grant No. PCIG14-GA-2013-631571 and from the DFG through 
the DIP Program (No. FO 703/2-1).
%\end{acknowledgements}

\end{document}